# Polaronic transport through molecular quantum dots: charging-induced NDR and rectifying behavior

Kamil Walczak

Institute of Physics, Adam Mickiewicz University
Umultowska 85, 61-614 Poznań, Poland

Here we study the polaronic transport through molecules weakly connected to metallic electrodes in the nonlinear response regime. Molecule itself is treated as a quantum dot with discrete energy levels, its connection to the electrodes is described within the wide-band approximation, while the charging is incorporated by means of the self-consistent potential. Nonperturbative computational scheme, used in this work, is based on the Green's function theory within the framework of mapping technique (GFT-MT). This method transforms the many-body electron-phonon interaction problem into a one-body multi-channel single-electron scattering problem with occupation of polaron levels calculated in a self-consistent way. In particular, three different phenomena as a result of charging in polaronic transport via discrete quantum states are discussed in detail: the suppression of the current at higher voltages, negative differential resistance (NDR effect), and rectification.

## 1. Introductory remarks

Electronic conduction through molecular junctions, composed of molecules sandwiched between the electrodes, is of great importance because of their potential as future electronic devices [1-10]. The current-voltage ($I-V$) spectra of such nanojunctions were measured experimentally and negative differential resistance (NDR effect) [11-15] as well as rectifying behavior [16-19] were reported. Suggested possible mechanisms for NDR involve charging and/or conformational changes [20-24], while the dominant factors in inducing rectification are some geometric asymmetry in the molecular junction and in the electrostatic potential spatial profile [25-28]. Anyway, transport characteristics are usually discussed in the context of simple tunneling through concrete energy levels (molecular orbitals).

Since molecules involved into the conduction process can be thermally activated to vibrations (phonon modes are excited), their transport properties should be strongly affected by electron-phonon ($e-ph$) interactions in the case when electron spends enough time on the molecule [29]. The contact time $\tau_c$ of the conduction electron with the molecule can be estimated by a straightforward generalization of the uncertainty principle

$$\tau_c \cong \frac{\hbar \beta L}{\Delta E_G}, \qquad (1)$$

where $\hbar$ is Planck constant, $\beta \sim 1$ Å$^{-1}$ [30] is the structure-dependent decay length of the electron transfer process, $L$ is the length of the molecular bridge, while $\Delta E_G$ is the excitation gap between the injection energy and the isolated bridge frontier orbital energy. The above relation (Eq.1) has important physical implications. For short bridges with large gaps ($\sigma$- bonded systems), the contact time $\tau_c \sim fs$ is far too short for significant vibronic coupling. For longer bridges with smaller gaps ($\pi$-type systems), the contact time $\tau_c \sim ps$ is of order of magnitude comparable to vibrational period [31]. In the second case, the vibronic coupling can be strong enough to lead to polaronic transport through molecular bridge, where the electronic virtual excitations of polaron states create conduction channels.

Indeed, the effects of the $e-ph$ coupling have been demonstrated in inelastic electron tunneling spectra of small molecules adsorbed on metallic surfaces [32-37]. On the other hand, because of the small sizes of molecular-scale devices also electron-electron ($e-e$) interactions between charge carriers are important in determining their transport properties. In particular, Coulomb blockade in single molecules weakly connected to the electrodes [38-40] and Kondo effect in single molecules with well-defined spin and charge states [39-42] have been experimentally observed. Both interaction effects are non-negligible at molecular scale, since Coulomb charging energies of single molecules are of the same order of magnitude of the relaxation energies induced by the $e-ph$ coupling.

This work is devoted to the question of polaronic transport through the molecules weakly connected to the electrodes in the nonlinear response regime. Molecule itself is treated as a quantum dot with discrete energy levels, the molecule-metal connections are described within the wide-band model, while the charging is incorporated by means of the mean-field approximation. The calculations are performed using nonperturbative computational scheme, based on Green's function theory and the so-called mapping technique (GFT-MT). This method transforms the many-body electron-phonon interaction problem into a single-electron many-channel scattering problem [43-49] with occupation of particular polaron levels calculated in a self-consistent way. The real advantage of the nonperturbative treatment is that it does not involve any restrictions on the model parameters, being well-justified in the boundary case of higher voltages. Here we show that NDR and rectification can occur also in the case of polaronic transport due to charging effects.

## 2. Theoretical formulation of the problem

Now we briefly outline out theoretical approach. Let us consider the simplest possible situation in which the molecular quantum dot is represented by one spin-degenerate electronic level coupled to a single vibrational mode (primary mode) while being also connected to two reservoirs of non-interacting electrons. The Hamiltonian of the whole system can be written in the form

$$H = \sum_{k \in \alpha} \varepsilon_k c_k^+ c_k + \sum_{k \in \alpha} \left( \gamma_k c_k^+ c_l + h.c. \right) + \varepsilon_l c_l^+ c_l + \Omega a^+ a - \lambda \left( a + a^+ \right) c_l^+ c_l. \qquad (2)$$

The first terms describes the left ($\alpha = L$) and right ($\alpha = R$) electrodes, the second term describes the tunnel connection between the molecule and two reservoirs, while the last three terms represent molecular part of the Hamiltonian. Here $\varepsilon_k$ and $\varepsilon_l$ are single-electron energies of electronic states in the reservoirs and on the molecule, $\gamma_k$ is the strength of the molecule-reservoir connection, $\Omega$ is the phonon energy, $\lambda$ is the electron-phonon interaction parameter. Furthermore, $c_k$, $c_l$, $a$ and their adjoints are annihilation and creation operators for the electrons in reservoirs and on the molecular level, and for the primary phonon, respectively.

The problem we are facing now is to solve a many-body problem with phonon emission and absorption when the electron tunnels through the molecule. To carry out the calculations, we apply the so-called polaron transformation, where the electron states into the molecule are expanded onto the direct product states composed of single-electron states and $m$-phonon Fock states

$$|l,m\rangle = d_l^+ \frac{(a^+)^m}{\sqrt{m!}}|0\rangle, \quad (3)$$

where electron state $|l\rangle$ is accompanied by $m$ phonons, and $|0\rangle$ denotes the vacuum state. Similarly, the electron states in all the $\alpha$-reservoirs can be expanded onto the states

$$|k,m\rangle = c_k^+ \frac{(a^+)^m}{\sqrt{m!}}|0\rangle. \quad (4)$$

Such procedure enables us to map the many-body electron-phonon interaction problem into a multi-channel single-electron scattering problem, as shown in Fig.1 and discussed elsewhere [43-49]. After eliminating the reservoir degrees of freedom, we can present the effective Hamiltonian of the reduced molecular system in the form

$$H_{eff} = \sum_{m,\alpha}\left(\varepsilon_l^m + \Sigma_\alpha^m\right)|l,m\rangle\langle l,m| - \sum_m \lambda^m \left(|l,m+1\rangle\langle l,m| + |l,m\rangle\langle l,m+1|\right), \quad (5)$$

where

$$\varepsilon_l^m = \varepsilon_l + U_{SCF} + m\Omega, \quad (6)$$

$$\lambda^m = \lambda\sqrt{m+1}, \quad (7)$$

while the particular levels are redefined with the help of the self-consistent potential $U_{SCF} = UQ_l^m$ in order to take into account the charging effects. The $U$-parameter represents the on-level Hubbard-type interaction constant, while the occupation of particular channels (polaron levels) can be computed similarly as in the case of the generalized Breit-Wigner formula [50]

$$Q_l^m = \frac{2}{\pi}\int_{-\infty}^{+\infty} d\varepsilon \frac{f_L^m \Gamma_L + f_R^m \Gamma_R}{4\left[\varepsilon - \varepsilon_l^m\right]^2 + \left[\Gamma_L + \Gamma_R\right]^2}, \quad (8)$$

where

$$f_\alpha^m = \left[\exp[\beta(\varepsilon + m\Omega - \mu_\alpha)] + 1\right]^{-1} \quad (9)$$

is the Fermi distribution function. Because of the fact that the potential $U_{SCF}$ is determined by the occupation $Q_l^m$, while $Q_l^m$ depends on $U_{SCF}$ – both quantities are recalculated in the self-consistent procedure until convergence. Index $m$ determines the statistical probability to excite the phonon state $|m\rangle$ at finite temperature $\theta$, and therefore the accessibility of particular conduction channels is determined by a weight factor

$$P_m = [1-\exp(-\beta\Omega)]\exp(-m\beta\Omega), \quad (10)$$

where $\beta = 1/(k_B\theta)$ and $k_B$ is Boltzmann constant.

Since we neglect all the nonequilibrium phonon effects (due to the assumed high energy relaxation rate) as well as dissipative processes (due to the assumed isolation from the influence of external surrounding), the electron energies are constrained by the conservation law

$$\varepsilon_{in} + m\Omega = \varepsilon_{out} + n\Omega, \qquad (11)$$

where $\varepsilon_{in}$ is the energy of the incoming electron with the initial amount of phonons $m$, while $\varepsilon_{out}$ is the energy of outgoing electron with the final amount of phonons $n$, respectively. In practice, the basis set is truncated to a finite number of possible excitations $m = m_{max}$ in the phonon modes because of the numerical efficiency. The size of the basis set strongly depends on: phonon energy $\Omega$, the temperature of the system under investigation $\theta$, and the strength of the electron-phonon coupling constant $\lambda$.

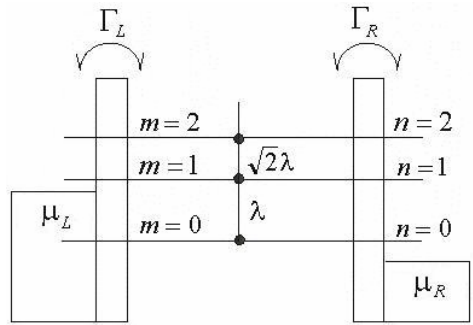

Figure 1: A schematic representation of inelastic scattering problem for the device composed of molecular quantum dot with single energy level connected to two metallic electrodes.

For simplicity, we adopt the wide-band approximation to treat both electrodes and the dephasing reservoir, where the self-energy and the so-called linewidth function are given through the relations

$$\Sigma_\alpha^m = -i\Gamma_\alpha^m / 2 \qquad (12)$$

and

$$\Gamma_\alpha^m = 2\pi \,|\, \gamma_k^m\,|^2 \rho_\alpha, \qquad (13)$$

respectively. Here $\gamma_k^m \equiv \gamma_\alpha$ is the energy and voltage independent parameter (by assumption) related to the strength of the effective connection between the $m$ th channel and the $\alpha$-reservoir characterized by constant density of states $\rho_\alpha$. Both electrodes are also identified with their electrochemical potentials

$$\mu_L = \varepsilon_F + \eta eV \qquad (14)$$

and

$$\mu_R = \varepsilon_F - (1-\eta)eV \qquad (15)$$

which are related to the Fermi energy level $\varepsilon_F$, while the voltage division factor is $\eta = 0.5$.

Now we proceed to analyze the problem of electron transfer between two reservoirs via discrete quantum state in the presence of phonons. An electron entering from the left hand side can suffer inelastic collisions by absorbing or emitting phonons before entering the right electrode. Such processes are presented graphically in Fig.1, where individual channels are indexed by the number of phonon quanta in the left ($m$) and right electrode ($n$), respectively.

Each of the mentioned processes is described by its own transmission probability, which can be written in the factorized form

$$T_{m,n}(\varepsilon) = \Gamma_L \Gamma_R |G_{m+1,n+1}(\varepsilon)|^2. \tag{16}$$

Such transmission function (Eq.16) is given through the matrix element of the molecular Green function of size $m_{max} \times m_{max}$ defined as

$$G(\varepsilon) = [J\varepsilon - H_{eff}]^{-1}. \tag{17}$$

Here $J$ stands for identity matrix, while $H_{eff}$ is the molecular Hamiltonian (Eq.5), while the effect of the connections with the $\alpha$-reservoirs is fully described by specifying self-energy corrections $\Sigma_\alpha$.

The total current flowing through the junction can be expressed in terms of transmission probability of the individual transition $T_{m,n}$ which connects incoming channel $m$ with outgoing channel $n$

$$I(V) = \frac{2e}{h} \int_{-\infty}^{+\infty} d\varepsilon \sum_{m,n} T_{m,n} \left[ P_m f_L^m (1 - f_R^n) - P_n f_R^n (1 - f_L^m) \right]. \tag{18}$$

The factor of 2 in Eq.18 accounts for the two spin orientations of conduction electrons. The elastic contribution to the current can be obtained from Eq.18 by imposing the constraint of elastic transitions, where $\varepsilon_{in} = \varepsilon_{out}$ or more precisely $n = m$. The differential conductance is then given by the derivative of the current with respect to voltage $G(V) = dI(V)/dV$, while the resistance $R = 1/G$.

## 3. Numerical results and discussion

To illustrate that method in a simple context, in this section we study a model of a molecule represented by a single relevant electronic level $\varepsilon_l$ coupled linearly to vibrational mode of energy $\Omega$ and symmetrically connected to the left ($L$) and right ($R$) wide-band metallic electrodes. This is a test case simple enough to analyze the essential physics of the problem in detail, while generalization to multilevel system with many different phonon quanta can be obtained straightforwardly. In our calculations we have used the following set of realistic parameters (given in eV): $\varepsilon_l = 0$ (the reference energy of the LUMO level), $\varepsilon_F = -1$, $\Omega = 1$, $\lambda = 0.5$, $\rho_L^{-1} = \rho_R^{-1} = 20$ (both electrodes are made of the same material), while the temperature of the system is set at $\theta = 300$ K ($\beta^{-1} = 0.025$). Maximum number of allowed phonons $m_{max} = 4$ is used to give fully converged results for all the chosen parameters.

Figure 2 presents nonlinear transport characteristics obtained for the symmetric anchoring case, i.e. the strength of the molecule-electrode connections is the same at both ends. The $I-V$ function reveals the well-known staircase-like structure, while for higher voltages we observe the suppression of the current due to the charging effect. When we neglect the vibronic coupling, only one current step positioned at $V_0 = 2|\varepsilon_l - \varepsilon_F|/e$ [= 2 V] is expected. However, in the presence of the strong $e-ph$ coupling, two current steps in the $I-V$ dependence (or equivalently two conductance peaks in the $G-V$ function) are observed due to the polaron formation. Using the formula for polaron energies it is possible to deduce the positions of the main conductance peak $V \cong V_0 - 2\lambda^2/(e\Omega)$ [= 1.5 V] and a one phonon side peak $V \cong V_0 + 2\Omega/e - 2\lambda^2/(e\Omega)$ [= 3.5 V]. The height of the second peak as associated with the first excited state of a polaron is much smaller than that of the first peak which corresponds to the polaron ground state.

Moreover, inclusion of the charging effects results in the significant suppression of the current for higher voltages. Here we also observe two current steps (two conductance peaks), but their positions are shifted in the direction to higher biases. Surprisingly, the charging-induced smoothing of the $I-V$ curve and the charging-induced broadening of the $G-V$ function are invisible. This conclusion stands in contradiction to the results obtained in the absence of phonons [50]. Besides, after the first current step the NDR effect is documented (differential conductance reaches negative values). Here we can formulate the following general conclusion: the higher is the value of the $U$-parameter, the stronger the NDR effect is observed. The important thing is to note that the NDR can not be generated by only one of two considered interactions, but it is combined effect of both polaron formation and charging.

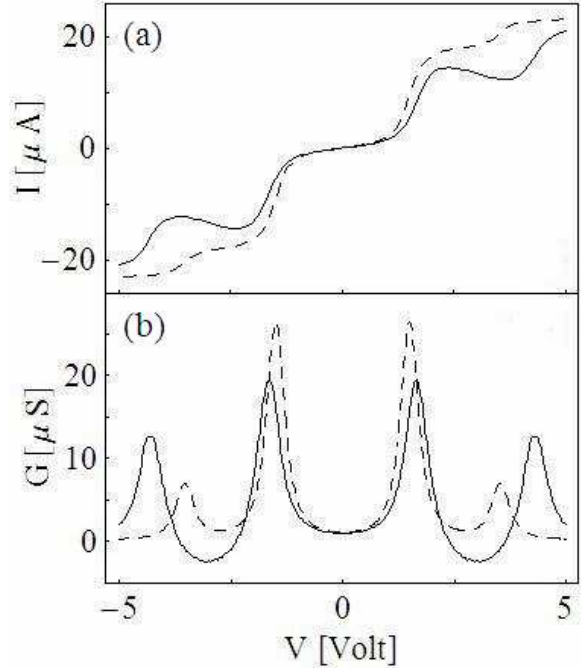

Figure 2: (a) Current-voltage ($I-V$) and (b) conductance-voltage ($G-V$) characteristics for molecular quantum dot symmetrically connected with two reservoirs ($\gamma_L = \gamma_R = 1$) for two different charging parameters: $U = 0$ (dashed lines) and $U = 2$ (solid lines). The other parameters of the model are given in the text.

In Fig.3 we plot transport characteristics for the asymmetric anchoring case, i.e. the strength of the molecule-electrode connections is different at both ends. This situation can be realized experimentally by adjusting the molecule-electrode bond length or by linking the molecule with two electrodes with the help of different anchoring groups. Our calculations indicate that both $I-V$ and $G-V$ dependences are symmetric in the absence of charging (for $U = 0$). The rectification effect, in which the magnitude of the junction current depends on the bias polarity, is observed for the case of non-zero charging energy parameter (for $U > 0$). Within our model, the diode-like behavior for higher voltages is due to a combined effect of asymmetric connections with the electrodes and charging itself. It is easy to draw the following general conclusion: asymmetry in transport characteristics increases with increasing the value of the $U$-parameter. To quantify the asymmetry of the $I-V$ curve, we plot the rectification ratio $RR(V) = |I(V)/I(-V)|$ in the inset of Fig.3. Here we show that for the chosen set of model parameters, the junction rectifies with factor $1 < RR < 2$.

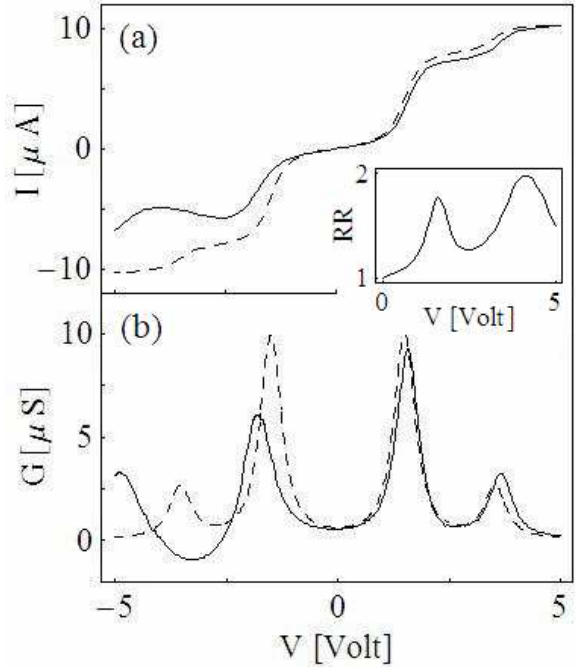

Figure 3: (a) Current-voltage ($I-V$) and (b) conductance-voltage ($G-V$) characteristics for molecular quantum dot asymmetrically connected with two reservoirs ($3\gamma_L = 1.5 = \gamma_R$) for two different charging parameters: $U = 0$ (dashed lines) and $U = 2$ (solid lines). The other parameters of the model are given in the text. The inset displays the rectification coefficient $RR$ as a function of bias voltage $V$.

In summary, we have presented a general method that can be used to study nonlinear transport properties of molecular devices in the presence of strong $e-ph$ coupling, where charging is taken into consideration via the self-consistent potential. Nonperturbative computational scheme is based on the Green's function theory within the framework of mapping technique (GFT-MT). This is an exact method to treat the $e-ph$ interactions, while the charging is included at the level of the mean-field approach. Our results show that transport is mediated via polaron propagation. In particular, the three different phenomena as a result of charging in polaronic transport through molecular quantum dots were documented: the suppression of the current at higher voltages, the NDR effect observed after the first current step in the $I-V$ dependence, and the rectification effect in the asymmetric-connection case.

Finally, it should be also mentioned that recently Galperin *et al.* have presented 'purely' polaron model, where all the charging effects are omitted [51]. They have suggested that polaronic mechanism can be responsible for NDR and hysteric/switching behavior in molecular junctions. In their model, the self-consistency is associated with the energy of the resonant level shifted by polaron formation that in turn depends on the electronic occupation in that level. Here, in contrast, we have described slightly different approach to polaronic transport, where the self-consistency is related to the energy of polaron level shifted by charging energy that in turn depends on the electronic occupation in that level.